\documentclass[twocolumn,showpacs,preprintnumbers,amsmath,amssymb]{revtex4}
\usepackage{graphicx}
\begin{document}
\title{Non-adiabatic coupling and adiabatic population transfer\\
in quantum molecular systems}
\author{Ignacio R. Sol\'a$^{1}$ and Vladimir S. Malinovsky$^{2}$}
\address{$^{1}$Department of Chemistry, Princeton University, 
Princeton, NJ 08544\\
$^{2}$  Michigan Center for Theoretical Physics 
$\&$ FOCUS Center, Department of Physics,
University of Michigan, Ann Arbor, MI 48109}
\begin{abstract} 
We show that a counter-intuitive pulse sequence 
leads to adiabatic passage between the vibrational levels
of three harmonic potentials through parallel dark 
states in adiabatic approximation. However, the adiabatic 
assumptions break down for very intense pulses and 
non-adiabatic couplings result in the population transfer
by light-induced potential shaping.
\end{abstract}
\pacs{33.80.-b, 42.50.-p}
\maketitle
\centerline{} 
Population transfer by shaping light induced potentials (LIP) is
a very robust scheme to transfer vibrational population between
different electronic potentials in diatomic 
molecules~\cite{newPRL,SolPRA00,us,MalCPH01,GarPRL98,RodPRA00}.
One of the main advantages
of the scheme is that the vibrational quantum number is
conserved during the overall process. Since the selectivity
of the transfer is guaranteed by a symmetry rule, the method is
not very sensitive to the exact position of the energy resonance, 
i.e. to the topological or energetic features of the potentials 
involved.

The simplest scenario of population transfer by
shaping LIP can be described in terms of
three electronic states (potential curves) sequentially
coupled by two laser pulses, and the vibrational population
is driven by two photon off resonant absorption.
Although there are several possible
scenarios for selective population transfer~\cite{newPRL,BanPRA01},
the first scheme proposed, called APLIP, involves a sequence
of two very strong (TW/cm$^2$) 
picosecond pulses applied in counter-intuitive order~\cite{GarPRL98}.
The pulse that couples the intermediate electronic state with the
final excited electronic state must precede the pulse that couples 
the ground electronic state with the intermediate one.
Then the system dynamics follows a characteristic pattern in the
vibrational basis representation, which is illustrated in
Fig.~1 for a test system of three harmonic oscillators with
the same force constant. (More details about the model are
presented below.) According to Fig.~1(a), the overall population
on the ground electronic state is rapidly transferred to the
final electronic state, while very few population is temporally
excited to the intermediate electronic state. 
So, in principal it is possible to suppose there is a dark state 
consisting of the initial and final potentials, similar to the 
population transfer through the dark state of the three-level 
system \cite{BeRMP98}. Looking into the vibrational populations 
in detail, it can be seen that the passage is mediated by 
substantial excitation of high energy vibrational levels in both 
initial and final potentials, Fig.1(b).

The scenario previously considered is very similar to the well known 
Stimulated Raman Adiabatic Passage (STIRAP) scheme~\cite{BeRMP98}. 
In STIRAP there are
three levels sequentially coupled by two fields working in 
counter-intuitive order. This arrangement of pulses prepares the
system initially in a dark state, $|\Phi_0\rangle$, 
which is an eigenstate of the
Hamiltonian dressed by the field in the adiabatic approximation. 
The dark state correlates at late times with 
the final level $|3\rangle$ and never overlaps with the intermediate
level. The population is transferred from $|1\rangle$ to $|3\rangle$
in a completely similar way to the overall electronic populations 
of Fig.~1(a).
\begin{figure}[h]
\includegraphics[height=7cm,width=6cm]{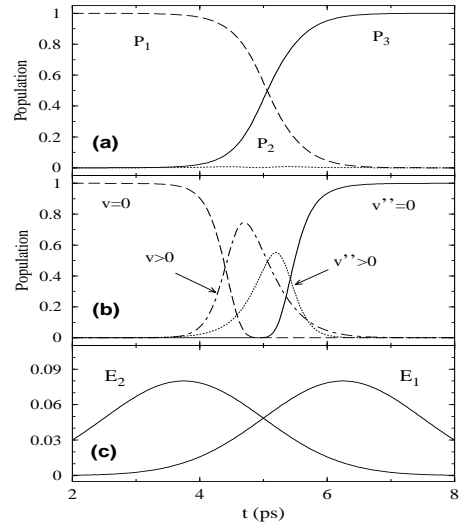}
\noindent 
\caption{Population dynamics in three symmetrically displaced
harmonic oscillators model. (a) Solid line shows the total population 
of final state,  dashed line shows the total population of ground 
state, and dotted line is the population of intermediate state.
(b) Population of vibrational levels, dashed line - zero vibrational 
level of ground state, dot-dashed line - population of the 
vibrational levels $v>0$ of ground state, solid line - population 
of zero vibrational level of final state, dotted line - population 
of the vibrational levels $v''>0$ of the final state. 
(c) Shape of the laser pulses.}
\end{figure}

Normally, the STIRAP scheme involves
nanosecond (or longer) pulses with intensities in the GW/cm$^2$
(or less). Therefore it is natural to ask if APLIP is a short time, 
strong field version of STIRAP (and applied to electronic states 
instead of vibrational or atomic levels). Indeed the
requirement of stronger laser fields can be understood
under the general assumptions of STIRAP. Since the adiabaticity
condition in STIRAP is usually expressed by the pulse area
relation, $\Omega \tau \gg 1$, where $\Omega$ is
the effective Rabi frequency of the pulses and $\tau$ their time
widths, any reduction in the pulse width must imply a corresponding
increase in the pulse amplitude. Nevertheless, in this note we
show that the population transfer in APLIP can not be reduced 
to the generalized description of STIRAP in multi-level system. 
In order to do so we show that the natural extension of STIRAP 
to the three potential curves scenario in the strong field limit 
predicts a different dynamical evolution than the one shown 
in Fig.~1.
\begin{figure}[t]
\centerline{\includegraphics[height=5cm,width=4cm]{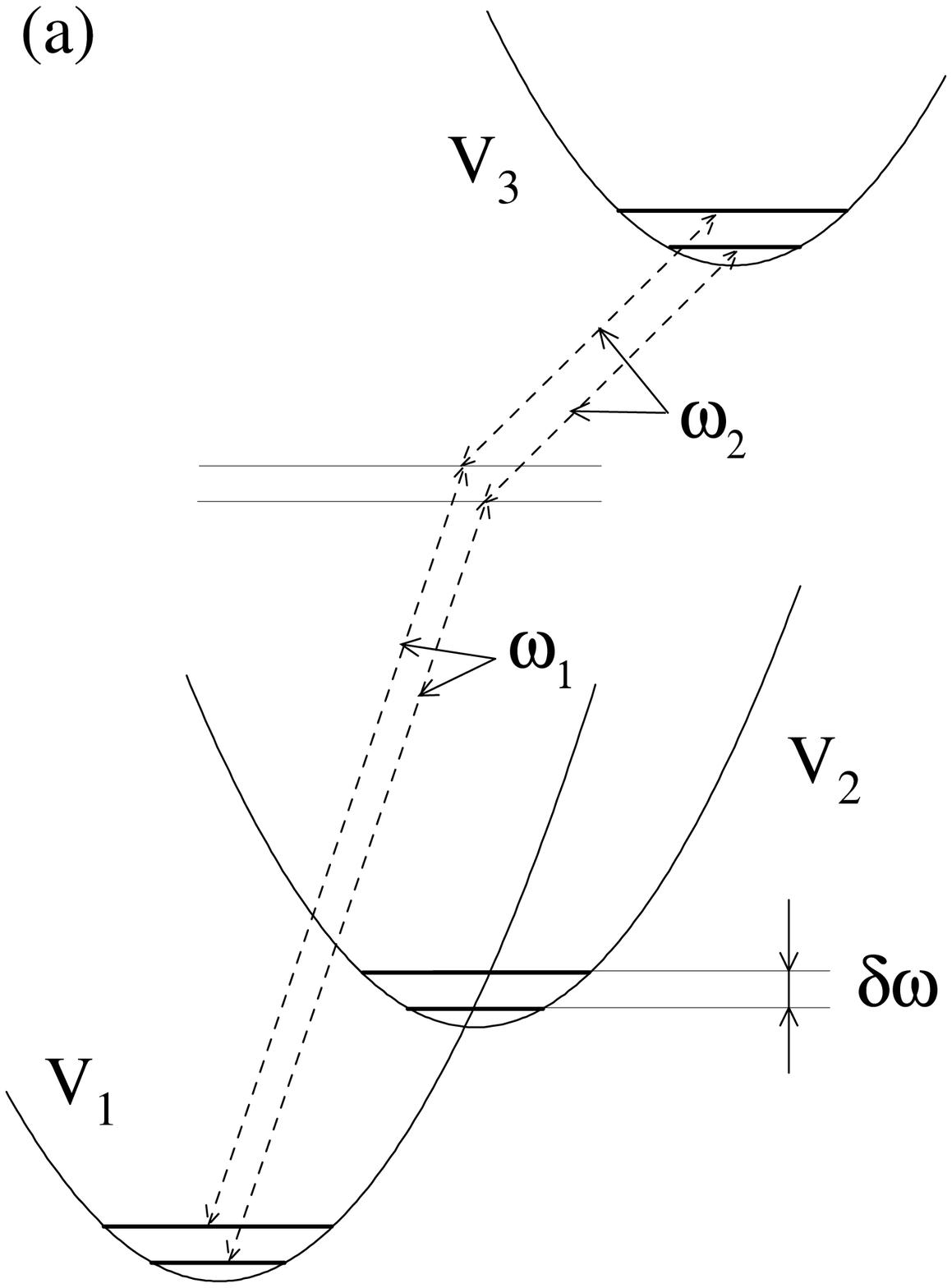}
\includegraphics[height=5cm,width=4cm]{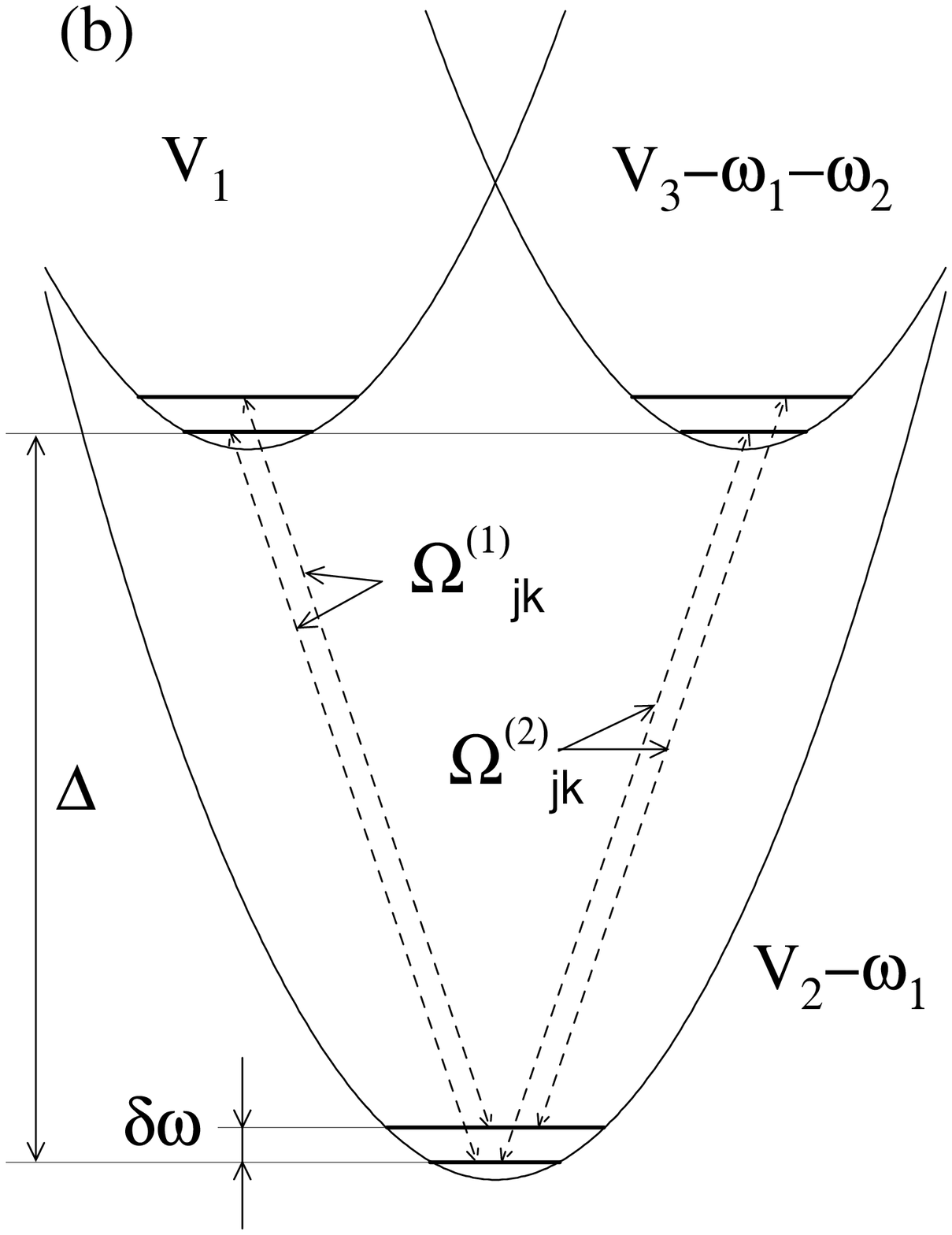}}
\noindent 
\vspace{0.5cm}
\caption{(a). Scheme of three symmetrically displaced
harmonic oscillators truncated to give two coupled
3-level ladder systems.
(b). Dressed potentials, $U_1=V_1$, $U_2=V_2-\omega_1$, 
and $U_3=V_3-\omega_1-\omega_2$.}
\end{figure}

Let us consider in detail the Hamiltonian of the system.
The Hamiltonian for three electronic potentials ($V_i(x)$, $i=1,2,3$) 
coupled by two laser fields ($E_1(t)$ and $E_2(t)$)
in the Born Oppenheimer and rotating wave approximation (RWA)
reads
\begin{widetext}
\begin{equation}
H^{RWA} =
\left( \begin{array}{ccc} T + U_1(x) &  - \frac{1}{2} \mu_{12}(x)E_1(t) 
& 0 \\ - \frac{1}{2} \mu_{12}(x)E_1(t) & T + U_2(x)  & - \frac{1}{2} 
\mu_{23}(x)E_2(t) \\ 0 & - \frac{1}{2} \mu_{23}(x)E_2(t) & T + U_3(x) \\
\end{array} \right) ,  
\label{eq:TDSERWA}
\end{equation}
\end{widetext}
\noindent
where $T$ is the kinetic energy operator and 
$U_i(x)$ are the diabatic dressed potentials ($U_1(x)=V_1(x)$,
$U_2(x) = V_2(x)-\omega_1$, $U_3(x)=V_3(x)-\omega_1
-\omega_2$, $\omega_i$ being the carrier frequencies of the lasers),
$\mu_{12}(x)$ and $\mu_{23}(x)$ are the dipole moments 
(atomic units are used throughout).
In this representation the global wave function of the system is
expanded in terms of electronic wave functions, $\{ \Xi_i(q;x) \}$,
and vibrational wave packets, $\{ \psi_i(x,t) \}$, where $q$ is an 
index representing the collective electron coordinates and $x$ is 
the vibrational coordinate. 

After expanding the vibrational wave packet in the set of 
eigenfunctions of the vibrational Hamiltonian for each potential,
\begin{equation}
\psi_{\alpha}(x,t) = \sum_j d_j^{(\alpha)}(t) \phi_j^{(\alpha)}(x) ,
\label{eq:psienlev}
\end{equation}
where $\phi_j^{(\alpha)}(x)$ is the j'th vibrational level in the
electronic state $\alpha$ (in the following we use
Greek letters to designate electronic states and Roman letters to 
designate vibrational levels), we arrive at the following time 
dependent Schr\"odinger equation (TDSE): 

\begin{equation}
\left\{
\begin{array}{ll}
&i\dot{d}^{(1)}_j = (\omega^{(1)}_j + \omega_1 + D_1^0 - D_2^0) d^{(1)}_j -
\sum_k \frac{\Omega^{(1)}_{jk}}{2} d^{(2)}_k \\
&i\dot{d}^{(2)}_j = - \sum_k \frac{\Omega^{(1)}_{jk}}{2} d^{(1)}_k +
(\omega^{(2)}_j - D_2^0) d^{(2)}_j - \sum_k \frac{\Omega^{(2)}_{jk}}{2} 
d^{(3)}_k \\
&i\dot{d}^{(3)}_j = - \sum_k \frac{\Omega^{(2)}_{jk}}{2} d^{(2)}_k +
(\omega^{(3)}_j -\omega_2 +D_3^0 - D_2^0) d^{(3)}_j  ,
\end{array} \right.
\label{eq:eq3ngen}
\end{equation}
\noindent
where $\omega^{(\alpha)}_j + D^0_{(\alpha)}$ is the eigenvalue 
corresponding to the $| \phi^{(\alpha)}_j\rangle$ eigenfunction, 
$\Omega^{(1)}_{ij} = E_1(t) \langle \phi^{(1)}_i \Xi_1
| \mu_1 | \phi^{(2)}_j \Xi_2 \rangle$ and $\Omega^{(2)}_{ij} = E_2(t)
\langle \phi^{(3)}_i \Xi_3 | \mu_2 | \phi^{(2)}_j \Xi_2 \rangle$
are the Rabi frequencies, and 
$D_{\alpha}^0$ are the potential zero energies.

In order to compare the dynamics of the general system with that of 
STIRAP, we establish a correspondence between the $3 \times N$ 
equations of motion (Eq.~(\ref{eq:eq3ngen})) and the equations for
$N$ $3$-level ladder systems, where both diabatic and adiabatic
states are known, corresponding to that of the STIRAP Hamiltonian.
In doing this connection we neglect all the contributions from
the continuum wave functions in all the electronic states. 
To simplify the notation we use a model of three symmetrically displaced
harmonic oscillators (Fig.~2) with the same force constant
(SDHO model), so that the separation between the
minima of the potentials is constant, $r_0^{(3)}-r_0^{(2)} =
r_0^{(2)}-r_0^{(1)} = R$. We assume that the equilibrium 
configurations of the excited potentials are displaced to larger
inter nuclei distances. Furthermore we  consider only
processes in two photon resonance. This allows us to define
a constant energy splitting, 
$\delta\omega = \omega_{n+1}^{(1)}-\omega_n^{(1)} =
\omega_{n+1}^{(3)}-\omega_n^{(3)}$, and a 
one photon detuning, $\Delta = D_2^0 -\omega_0^{(1)} -\omega_1 = 
(\omega_2+D_2^0) -\omega_0^{(3)} -D_3^0$. The TDSE is:
\begin{equation}
\left\{
\begin{array}{ll}
&i\dot{d}^{(1)}_j = (j\delta\omega - \Delta) d^{(1)}_j - \sum_k
\frac{\Omega^{(1)}_{jk}}{2} d^{(2)}_k \\
&i\dot{d}^{(2)}_j = - \sum_k \frac{\Omega^{(1)}_{jk}}{2} d^{(1)}_k +
(j\delta\omega) d^{(2)}_j - \sum_k \frac{\Omega^{(2)}_{jk}}{2} d^{(3)}_k \\
&i\dot{d}^{(3)}_j = - \sum_k \frac{\Omega^{(2)}_{jk}}{2} d^{(2)}_k +
(j\delta\omega - \Delta) d^{(3)}_j .
\end{array} \right.
\label{eq:eq3ndho}
\end{equation}

Finally we invoke the Condon approximation.
Therefore, the state to state Rabi frequencies can be expressed as
$\Omega^{(1)}_{ij} = p^{12}_{ij} \mu_1 E_1(t)$ and
$\Omega^{(2)}_{ij} = p^{23}_{ij} \mu_2 E_2(t)$, where we
have defined the geometric Franck-Condon parameters
$p_{ij}^{\alpha\beta}=\langle\phi_i^{(\alpha)} | \phi_j^{(\beta)}\rangle$.
This is the model used to obtain the results shown in Fig.~1,
where we have chosen $\delta\omega = 7\cdot 10^{-4}$ a.u., 
$\Delta = 0.015$ a.u., $\mu_1 = \mu_2 = 1$,
$E_i(t) = E_0 S(t)$, with $E_0 = 0.08$ a.u. and $S(t)$ being a Gaussian
envelope function.
Both pulses have the same envelope function, with width $\sigma = 2.5$ ps 
and they are time delayed, so that the second pulse precedes in $2.5$ ps
the first pulse. 
To obtain the numerical results shown in Fig.~1 we have numerically
integrated the TDSE based on a grid discretization of the
Hamiltonian in Eq.~(\ref{eq:TDSERWA}) and not on the discrete basis
representation of Eq.~(\ref{eq:eq3ndho}). Details of the numerical
propagator are given in reference~\cite{SolPRA00}.

In order to go further in the comparison between the set of 
Eqs.~(\ref{eq:eq3ndho}) and those of STIRAP, we change the
representation from the diabatic basis $\{\phi^{(1)}_n, \phi^{(2)}_n, 
\phi^{(3)}_n \}$ to the adiabatic basis of the 3-levels ladder
system for each $n$ sub-system, $\{ \Phi^{(n)}_+, \Phi^{(n)}_0,
\Phi^{(n)}_- \}$, using the block diagonal rotation matrix
$\widehat{R}= \widehat{R}_1 \ldots \oplus \widehat{R}_n \ldots 
\oplus \widehat{R}_N$
with 
\begin{equation}
\widehat{R}_n = \left( \begin{array}{ccc}
\sin\varphi_n \sin\theta_n & \cos\theta_n & \cos\varphi_n \sin\theta_n \\
\, & \, & \, \\
\cos\varphi_n & 0 & -\sin\varphi_n \\
\, & \, & \, \\
\sin\varphi_n \cos\theta_n & -\sin\theta_n & \cos\varphi_n \cos\theta_n \\
\end{array} \right) ,  \label{eq:Rtransf}
\end{equation}
where the angles are defined by 
$\tan\theta_n = \Omega^{(1)}_{nn} / \Omega^{(2)}_{nn}$ and
$\tan (2\varphi_n) = 2 \left( \sqrt{(\Omega^{(1)}_{nn})^2+
(\Omega^{(2)}_{nn})^2} \right) / \Delta$.

In the new representation, the dynamics of the system is followed
by the amplitude coefficients $\{ a^{(n)}_+, a^{(n)}_0, a^{(n)}_- \}$,
which are delocalized and follow the electronic transitions between 
the potentials at different moments of time. The new
basis is {\em quasi-adiabatic} since
the transformation matrix $\widehat{R}$ diagonalizes only each
$3 \times 3$ sub-system matrices, while there remain couplings
between different sub-systems. 

To illustrate the nature of these
couplings, we detail the analysis for a system of 6-levels
obtained by truncating the expansion of the wave function
(Eq.~(\ref{eq:psienlev})) to only the first two vibrational levels.
We consider these levels as belonging to two coupled 3-level
ladder systems, whose Hamiltonian is
\begin{widetext}
\begin{equation}
H = 
\left( \begin{array}{cc}
\widehat{H}_{11} & \widehat{H}_{12} \\ 
\, & \, \\
\widehat{H}_{21} & \widehat{H}_{22} \\ 
\end{array} \right) =
- \frac{1}{2} \left( \begin{array}{ccccccc}
0 & \Omega^{(1)}_{11} & 0 & 0 & \Omega^{(1)}_{12} & 0 \\
\, & \, & \, & \, & \, &\, \\
\Omega^{(1)}_{11} & 2\Delta & \Omega^{(2)}_{11} &
\Omega^{(1)}_{12} & 0 & \Omega^{(2)}_{12} \\
\, & \, & \, & \, & \, &\, \\
0 & \Omega^{(2)}_{11} & 0 & 0 & \Omega^{(2)}_{12} & 0 \\
\, & \, & \, & \, & \, &\, \\
0 & \Omega^{(1)}_{21} & 0 & - 2\delta\omega & \Omega^{(1)}_{22} & 0 \\
\, & \, & \, & \, & \, &\, \\
\Omega^{(1)}_{21} & 0 &\Omega^{(2)}_{21} & \Omega^{(1)}_{22} &
- 2(\delta\omega - \Delta) & \Omega^{(2)}_{22} \\
\, & \, & \, & \, & \, &\, \\
0 & \Omega^{(2)}_{21} & 0 & 0 & \Omega^{(2)}_{22} & - 2\delta\omega \\
 \end{array} \right) ,
\label{eq:H6l}
\end{equation}
\end{widetext}
\noindent
and we next change the representation using
$\widehat{R} = \widehat{R}_1\oplus\widehat{R}_2$ ($\widehat{R}_n$ is
given by Eq.~(\ref{eq:Rtransf})).
We obtain the quasi-adiabatic Hamiltonian
\begin{equation}
\begin{array}{lll}
\widehat{H}^{CD} & = \left( \begin{array}{cc}
\widehat{R}_1^{-1} & 0 \\ 0 & \widehat{R}_2^{-1} \\
\end{array} \right) \left( \begin{array}{cc}
\widehat{H}_{11} & \widehat{V}_{12} \\ \widehat{V}_{21} &
\widehat{H}_{22} + \delta\omega \widehat{I} \\
\end{array} \right) \left( \begin{array}{cc}
 \widehat{R}_1 & 0 \\ 0 & \widehat{R}_2 \\
\end{array} \right) \\ &\, \\
\\ & = \left( \begin{array}{cc}
 \widehat{H}^{(1)} & \widehat{R}_1^{-1} \widehat{V}_{12} \widehat{R}_2 \\
\widehat{R}_2^{-1} \widehat{H}_{21} \widehat{R}_1
& \widehat{H}^{(2)} + \delta\omega \widehat{I} \\
\end{array} \right) .
\end{array}
\end{equation}

In this representation each sub-system Hamiltonian $\widehat{H}^{(n)}$ 
is diagonal. Neglecting
the coupling between sub-systems, $\widehat{H}^{QA}_{12} = 
\widehat{R}_1^{-1} \widehat{V}_{12} \widehat{R}_2 \approx 0$, and using
a counter-intuitive sequence of pulses, the Hamiltonian
has two parallel (independent) dark states as in STIRAP.
Therefore, the differences between APLIP and STIRAP must come from
the inter sub-system couplings.  %,
Substituting $\cos (\theta_n) = \Omega^{(2)}_{nn} / \Omega^{(e)}_{nn}$
and $\sin (\theta_n) = \Omega^{(1)}_{nn} / \Omega^{(e)}_{nn}$ where
$\Omega^{(e)}_{nn} = \left( (\Omega^{(1)}_{nn})^2 +
(\Omega^{(2)}_{nn})^2 \right)^{\frac{1}{2}}$, we obtain, for our
reduced 6-level Hamiltonian:
%, which in our reduced 6-level Hamiltonian take the form
\begin{widetext}
\begin{equation}
\widehat{H}^{QA}_{12} = (\widehat{H}^{QA}_{21})^\dagger =
{\footnotesize
\left( \begin{array}{ccc}
{ \begin{array}{l}
\!\!\!\frac{\sin\varphi_1\cos\varphi_2}{\Omega^{(e)}_{11}} 
(\Omega^{(1)}_{11} \Omega^{(1)}_{12} +\Omega^{(2)}_{11} \Omega^{(2)}_{12} ) \\
+ \frac{\cos\varphi_1\sin\varphi_2}{\Omega^{(e)}_{22}} (\Omega^{(1)}_{22}
\Omega^{(1)}_{12} + \Omega^{(2)}_{22} \Omega^{(2)}_{12} ) \end{array}} &
\!\!\!-\frac{\cos\varphi_1}{\Omega^{(e)}_{22}} ( \Omega^{(1)}_{22} 
\Omega^{(2)}_{12} - \Omega^{(2)}_{22} \Omega^{(1)}_{12} )  &
{ \begin{array}{l}
\!\!\!- \frac{\sin\varphi_1\sin\varphi_2}{\Omega^{(e)}_{11}} 
(\Omega^{(1)}_{11} \Omega^{(1)}_{12} +\Omega^{(2)}_{11} \Omega^{(2)}_{12} ) \\
+ \frac{\cos\varphi_1\cos\varphi_2}{\Omega^{(e)}_{22}} (\Omega^{(1)}_{22}
\Omega^{(1)}_{12} - \Omega^{(2)}_{22} \Omega^{(2)}_{12} ) \end{array}} \\ \\
\!\!\!\!-\frac{\cos\varphi_2}{\Omega^{(e)}_{11}} ( \Omega^{(1)}_{11} 
\Omega^{(2)}_{12} \! - \Omega^{(2)}_{11} \Omega^{(1)}_{12} ) & \!\!\!\! 0 &
\!\!\!\! \frac{\sin\varphi_2}{\Omega^{(e)}_{11}} ( \Omega^{(1)}_{11} 
\Omega^{(2)}_{12} - \Omega^{(2)}_{11} \Omega^{(1)}_{12} ) \\ \\
{ \begin{array}{l}
\!\!\!\!\frac{\cos\varphi_1\cos\varphi_2}{\Omega^{(e)}_{11}} 
(\Omega^{(1)}_{11} \Omega^{(1)}_{12} +\Omega^{(2)}_{11} \Omega^{(2)}_{12} ) \\
\!- \frac{\sin\varphi_1\sin\varphi_2}{\Omega^{(e)}_{22}} (\Omega^{(1)}_{22}
\Omega^{(1)}_{12} + \Omega^{(2)}_{22} \Omega^{(2)}_{12} ) \end{array} } &
\!\!\!\!\frac{\sin\varphi_2}{\Omega^{(e)}_{22}} ( \Omega^{(1)}_{22} 
\Omega^{(1)}_{12} - \Omega^{(2)}_{22} \Omega^{(1)}_{12} ) &
{ \begin{array}{l}
\!\!\!\!-\frac{\cos\varphi_1\sin\varphi_2}{\Omega^{(e)}_{11}}
(\Omega^{(1)}_{11} \Omega^{(1)}_{12} + \Omega^{(2)}_{11} \Omega^{(2)}_{12} ) \\
\!- \frac{\sin\varphi_1\cos\varphi_2}{\Omega^{(e)}_{22}} (\Omega^{(1)}_{22}
\Omega^{(1)}_{12} + \Omega^{(2)}_{22} \Omega^{(2)}_{12} ) \end{array} }
\end{array} \!\!\!\right) } .
\end{equation}
\end{widetext}
\noindent

Now let us concentrate in the case when both Rabi frequencies are 
equal, $\mu_{12} A_1 = \mu_{23} A_2 = \Omega_0$.
We consider the initial state to be the ground vibrational level, 
$|\phi_0^{(1)}\rangle$ which initially correlates with 
$|\Phi_0^{(1)}\rangle$.
Then the only non zero couplings between this state and any other
quasi-adiabatic state are
\begin{widetext}
\begin{equation}
\langle\Phi_+^{(2)}|H^{QA}_{21}|\Phi_0^{(1)}\rangle = -\cos{\varphi_2}
(p_{11}^{12}p_{21}^{23}-p_{21}^{12}p_{11}^{23})
\frac{\Omega_0 S_1(t)S_2(t)}
{\sqrt{(\Omega^{(1)}_{11}(t))^2+(\Omega^{(2)}_{11}(t))^2}} \ \  ,
\label{eq:ac0H6la}
\end{equation}
and
\begin{equation}
 \langle\Phi_-^{(2)}|H^{QA}_{21}|\Phi_0^{(1)}\rangle = \sin{\varphi_2}
(p_{11}^{12}p_{21}^{23}-p_{21}^{12}p_{11}^{23})
\frac{\Omega_0 S_1(t)S_2(t)}
{\sqrt{(\Omega^{(1)}_{11}(t))^2+(\Omega^{(2)}_{11}(t))^2}} \ \  .
\label{eq:ac0H6lb}
\end{equation}
\end{widetext}
Equations~(\ref{eq:ac0H6la}), and~(\ref{eq:ac0H6lb}) separate the 
molecular contribution (geometrical
factors) from the pulse contribution (depending on the pulse amplitude).
However it can be seen that the APLIP dynamic behavior cannot stem
from these terms. For instance, in the model of
symmetrically displaced harmonic oscillators 
the symmetry of the system imposes that
$p_{12}^{12} = p_{21}^{23}$ and $p_{11}^{23} = p_{11}^{12}$.
Therefore the quasi-adiabatic initially populated state, 
$|\Phi_0^{(1)}\rangle$,
is not coupled with the rest of the system. So, this state is indeed the
adiabatic state whose evolution reproduces exactly the STIRAP behavior
independently from the detuning or intensity of the pulses.

Increasing the number of vibrational levels in the expansion 
(Eq.~(\ref{eq:psienlev})) does not introduce any different type of 
couplings and the same conclusions apply.
It is very simple to generalize the expression for the general 
$3N$-level system, in this case, the Hamiltonian matrix terms are:
\begin{widetext}
\begin{equation}
\begin{array}{ll}
&\langle \Phi_0^{(n)} | H^{QA} | \Phi_0^{(m)} \rangle = 0 \\
&\langle \Phi_0^{(n)} | H^{QA} | \Phi_+^{(m)} \rangle =- \Omega_0 
\cos{\varphi_m}
( p_{nn}^{12}p_{nm}^{23} - p_{nm}^{12}p_{nn}^{23} )
\frac{S_1(t) S_2(t)}{\sqrt{(p_{nn}^{12}S_1(t))^2+(p_{nn}^{23}S_2(t))^2}} \\
&\langle \Phi_0^{(n)} | H^{QA} | \Phi_-^{(m)} \rangle =
\Omega_0 \sin{\varphi_m}
( p_{nn}^{12}p_{nm}^{23} - p_{nm}^{12}p_{nn}^{23} )
\frac{S_1(t) S_2(t)}{\sqrt{(p_{nn}^{12}S_1(t))^2+(p_{nn}^{23}S_2(t))^2}} .
\end{array}
\end{equation}
\end{widetext}

For the SDHO model, the geometrical symmetry requires that:
$p_{11}^{12} = \langle\phi_1^{(1)} |\phi_1^{(2)}\rangle
= \langle\phi_1^{(2)} |\phi_1^{(3)}\rangle = p_{11}^{23}$,
and $p_{12}^{12} = \langle\phi_1^{(1)} |\phi_2^{(2)}\rangle
= \langle\phi_2^{(2)} |\phi_1^{(3)}\rangle = p_{21}^{23}$. Once
again, this makes any couplings involving the $|\Phi_0^{(n)} \rangle$ 
states identically zero. The system exhibits $N$ parallel 
(independent) trapped states.
Therefore, if the initial wave function is a
coherent superposition of vibrational levels of the ground
electronic state,
$|\Psi(x,0) \rangle = \sum_{n=0}^{N} a_n(0) |\phi^{(1)}_n(x)\rangle$,
a sequence of two coherent counter-intuitive pulses will read this
wave function in the adiabatic representation as
$|\Psi(x,0)\rangle = \sum_{n=0}^{N} a_n(0) |\Phi^{(n)}_0 (x,0)\rangle$,
that will adiabatically passage at final times to
$|\Psi(x,T) \rangle = \sum_{n=0}^{N} a_n(0) \exp{\left( -i(n-1)
\delta\omega T \right)} |\phi^{(3)}_n(x)\rangle$
(where the phase factor only depends on the eigenvalues of the
dressed states $|\Phi^{(n)}_0\rangle$, i.e. 
the geometrical or Berry phase is zero).
At each instant of time only the initial levels
$|\phi^{(1)}_n\rangle$ and the corresponding
$|\phi^{(3)}_n\rangle$ levels are populated, the final population
in $|\phi^{(3)}_n\rangle$ being completely determined by
the initial population, $|a_n(0)|^2$.
The same conclusion applies for the final probabilities if
the initial state is an incoherent sum of vibrational levels.
The dynamics follows as $N$ independent STIRAP type systems and does
not reproduce the characteristic features of APLIP.

This result is in flagrant contradiction with the numerical evidence
(direct solution of the TDSE with the Hamiltonian of Eq.(1)), in which
many of high energy vibrational levels of the ground and final 
potentials are considerably populated (Fig.~1(b)). Although the analytic 
proof derived in this paper is valid for a specific ``ideal''
system, the SDHO model, of course any small asymmetry in the model
cannot explain the clear difference in behavior of STIRAP and APLIP.
The difference, therefore, must stem from a different source.
The validity of the quasi-adiabatic Hamiltonian rests upon the
adiabatic approximation, that is, the neglect of the contribution
from all the terms coming from $\widehat{R}^{-1}\dot{
\widehat{R}} = \widehat{R}_1^{-1}\dot{\widehat{R}}_1 \oplus
\widehat{R}_2^{-1}\dot{\widehat{R}}_2 \oplus \ldots \oplus
\widehat{R}_N^{-1}\dot{\widehat{R}}_N$, that couple states
$\Phi_0^{(n)}$ and $\Phi_{\pm}^{(n)}$ belonging to the same sub-system.
In STIRAP, the adiabatic condition, $\Omega_0 \tau \gg 1$ guarantees 
that these terms can be neglected. Surprisingly enough, for very
large $\Omega_0$ and detuning, the APLIP dynamics shows that
the contribution of these terms cannot be neglected. This is 
in agreement with the numerical results obtained for population transfer
by shaping LIPs in more general scenarios~\cite{us,RodPRA00}.

I.R.S. gratefully acknowledges support from the Secretar\'{\i}a de 
Estado de Educaci\'on y Universidades (Spanish Government).

\end{document}